\documentclass{article}

\usepackage{microtype}
\usepackage{graphicx}
\usepackage{booktabs}
\usepackage{array}
\usepackage{placeins}  % \FloatBarrier: appendix floats stay in their section
\graphicspath{{figures/}}

\usepackage{hyperref}

\usepackage[preprint]{icml2026}
\usepackage{amsmath}
\usepackage{amssymb}
\usepackage{amsfonts}
\usepackage{nicefrac}
\usepackage[capitalize,noabbrev]{cleveref}
\usepackage{subcaption}

\usepackage{dblfloatfix}

\newcommand{\Ztwo}{100}  % results/heldout/h2.json:headline.Z_r1.detection
\newcommand{\Xprime}{100}  % results/heldout/h2.json:headline.X_prime_classifier.detection
\newcommand{\TauRone}{0.629}  % results/heldout/thresholds.json:r1.threshold
\newcommand{\TauClf}{0.0}  % results/heldout/thresholds.json:classifier.fits.co-location.threshold
\newcommand{\FprRoneDecode}{0.6}  % results/heldout/h2.json:headline.r1_test_fpr.serving-decode
\newcommand{\FprRoneIdle}{0.1}  % results/heldout/h2.json:headline.r1_test_fpr.idle
\newcommand{\ExpDecodeH}{0.475}  % results/monitor/false_alarms.json:backgrounds.declared inference.r1.exposure_h
\newcommand{\FRoneDecode}{0.0}  % results/monitor/false_alarms.json:backgrounds.declared inference.r1.F_per_h
\newcommand{\FupRoneDecode}{6.3}  % results/monitor/false_alarms.json:backgrounds.declared inference.r1.F_upper95_poisson_model
\newcommand{\FClfDecode}{6.3}  % results/monitor/false_alarms.json:backgrounds.declared inference.classifier.F_per_h
\newcommand{\AClfDecode}{0.010}  % results/monitor/false_alarms.json:backgrounds.declared inference.classifier.A
\newcommand{\ARoneBusy}{0.998}  % results/monitor/false_alarms.json:backgrounds.non-AI busy.r1.A
\newcommand{\EpDelayRone}{3.0}  % results/monitor/episodes.json:detectors.r1.pooled_median_delay_s
\newcommand{\LminRone}{5.1}  % results/monitor/episodes.json:detectors.r1.Lmin_s
\newcommand{\GkHardPrefill}{100}  % results/gk/t5.json:forests.hardened.gate_inference.by_condition.serving-prefill.flag_rate
\newcommand{\GkGenPrefill}{2.4}  % results/gk/t5.json:forests.genuine.gate_inference.by_condition.serving-prefill.flag_rate
\newcommand{\GkGenMiss}{0}  % results/gk/t5.json:forests.genuine.co-location.per_run.panel-mixture-concurrent-0p5.detection
\newcommand{\GPU}{H200 NVL}  % paper/setup.json:gpu.name
\newcommand{\Probe}{Micsig RCP120}  % paper/setup.json:probe.model
\newcommand{\ProbeSens}{50}  % paper/setup.json:probe.sensitivity_mV_per_A
\newcommand{\ProbeLo}{34}  % paper/setup.json:probe.band_lo_hz
\newcommand{\ProbeHi}{30}  % paper/setup.json:probe.band_hi_mhz
\newcommand{\Fs}{199,808}  % paper/setup.json:sample_rate_sps
\newcommand{\EnvFrame}{2}  % paper/setup.json:envelope_frame_ms
\newcommand{\ModelParams}{168}  % paper/setup.json:model.params_millions
\newcommand{\ModelLayers}{8}  % paper/setup.json:model.n_layer
\newcommand{\ModelWidth}{1024}  % paper/setup.json:model.d_model
\newcommand{\RhoLo}{0.25}  % paper/setup.json:colocation_share.low
\newcommand{\RhoMid}{0.5}  % paper/setup.json:colocation_share.mid
\newcommand{\RhoHi}{0.75}  % paper/setup.json:colocation_share.high
\newcommand{\DefaultDecision}{0.5}  % paper/setup.json:default_decision
\newcommand{\GkNFeat}{30}  % paper/setup.json:gk.n_features
\newcommand{\GkRate}{9.765625}  % paper/setup.json:gk.rate_msps
\newcommand{\GkRateApprox}{10}  % paper/setup.json:gk.rate_msps
\newcommand{\GkDecimated}{200}  % paper/setup.json:gk.decimated_ksps
\newcommand{\GkFone}{0.955}  % paper/setup.json:gk.macro_f1
\newcommand{\GkAcc}{0.97}  % paper/setup.json:gk.accuracy
\newcommand{\GkAliasKhz}{10}  % paper/setup.json:gk.alias_khz
\newcommand{\NCaptures}{62}  % results/windows/manifest.json:n_captures
\newcommand{\CorpusHours}{23.4}  % results/windows/manifest.json:captures
\newcommand{\Win}{5}  % results/windows/manifest.json:window_s
\newcommand{\HopSnap}{5}  % results/windows/manifest.json:hops_s.0
\newcommand{\HopMon}{1}  % results/windows/manifest.json:hops_s.1
\newcommand{\CorpusCommit}{73ffebe}  % results/windows/manifest.json:corpus_commit
\newcommand{\NFeat}{25}  % results/replication/t3.json:n_features
\newcommand{\NRepTraces}{300}  % results/replication/t3.json:n_traces
\newcommand{\SoneFone}{0.933}  % results/replication/t3.json:two_stage.macro_f1
\newcommand{\Alpha}{0.05}  % results/heldout/h2.json:alpha
\newcommand{\FracCalZero}{98.8}  % results/heldout/thresholds.json:classifier.fits.co-location.frac_calibration_zero
\newcommand{\RoneMinGroup}{99.9}  % results/heldout/h2.json:r1.detection
\newcommand{\ClfJitterWorst}{76}  % results/heldout/h2.json:classifier.jitter.detection.all.per_run
\newcommand{\GateTrain}{80}  % results/gk/t5.json:forests.hardened.gate.training_min
\newcommand{\GateInf}{20}  % results/gk/t5.json:forests.hardened.gate.inference_max
\newcommand{\Rearm}{5}  % results/monitor/false_alarms.json:rule.rearm
\newcommand{\ExpIdleH}{8.35}  % results/monitor/false_alarms.json:backgrounds.idle.r1.exposure_h
\newcommand{\ExpDeclH}{8.8}  % results/monitor/false_alarms.json:backgrounds
\newcommand{\EpTotal}{192}  % results/monitor/episodes.json:detectors.r1.cells
\newcommand{\EpMinDur}{5.1}  % results/monitor/episodes.json:detectors.r1.cells
\newcommand{\EpMaxDur}{240}  % results/monitor/episodes.json:detectors.r1.cells
\newcommand{\Grace}{5}  % results/monitor/episodes.json:rule.grace_s
\newcommand{\EpTruncated}{4}  % results/monitor/episodes.json:detectors.r1.counts.truncated
\newcommand{\EpScored}{188}  % results/monitor/episodes.json:detectors.r1.cells
\newcommand{\LminLevel}{0.5}  % results/monitor/episodes.json:rule.lmin_level
\newcommand{\TfourMisCoPooled}{0}  % results/robustness/t4.json:misspecified_declaration.pooled.h2.r1.co-location.pooled
\newcommand{\TfourMisCoDeclared}{100}  % results/robustness/t4.json:misspecified_declaration.declared.h2.r1.co-location.pooled
\newcommand{\GkNTraces}{1610}  % results/gk/t5.json:forests.hardened.n_fit_traces
\newcommand{\GkOwnFone}{0.891}  % results/gk/t5.json:own_check.macro_f1_mean
\newcommand{\GkOwnFoneLo}{0.845}  % results/gk/t5.json:own_check.macro_f1_range.0
\newcommand{\GkOwnFoneHi}{0.945}  % results/gk/t5.json:own_check.macro_f1_range.1
\newcommand{\GkOwnAcc}{0.908}  % results/gk/t5.json:own_check.accuracy_mean
\newcommand{\GkShuffle}{0.306}  % results/gk/t5.json:own_check.negative_control.macro_f1
\newcommand{\GapGuard}{0.2}  % results/monitor/episodes.json:rule.gap_guard
\newcommand{\ARoneDecode}{0}  % results/monitor/false_alarms.json:backgrounds.declared inference.r1.A
\newcommand{\XzeroPrime}{0.6}  % results/heldout/operating_point.json:replica.tau_alpha.detection_per_run.panel-mixture-concurrent-0p5
\newcommand{\RepPrefill}{100}  % results/heldout/operating_point.json:replica.tau_alpha.fpr.development.serving-prefill
\newcommand{\ClfDevPrefill}{4}  % results/heldout/operating_point.json:classifier.tau_alpha.fpr.development.serving-prefill
\newcommand{\EpPreRone}{47}  % results/monitor/episodes.json:detectors.r1.counts.pre-alarmed
\newcommand{\GkTraceS}{5}  % paper/setup.json:gk.trace_s
\newcommand{\RtWinLo}{5}  % paper/setup.json:rt_window_s.0
\newcommand{\RtWinHi}{60}  % paper/setup.json:rt_window_s.1
\newcommand{\ChainBelowBand}{99.3}  % paper/setup.json:chain_check.train_var_frac_below_20khz
\newcommand{\ChainBandKhz}{20}  % paper/setup.json:chain_check.band_khz
\newcommand{\ChainEnvPct}{0.3}  % paper/setup.json:chain_check.envelope_match_pct_vs_2msps
\newcommand{\FupClfDecode}{16.3}  % results/monitor/false_alarms.json:backgrounds.declared inference.classifier.F_upper95_poisson_model
\newcommand{\RonePrefillHoldout}{100}  % results/declaration/heldout.json:results.6.check4.per_run.panel-serving-prefill.fp_rate
\newcommand{\TauDecOnlyPooled}{0.786}  % results/heldout/r1_scores.json:thresholds.decode_only.pooled calibration
\newcommand{\TauDecOnlyDecode}{0.569}  % results/heldout/r1_scores.json:thresholds.decode_only.decode calibration
\newcommand{\RoneAucDecode}{0.994}  % figures/f7_r1.json:annotated.r1.auc_unpaced_vs_test_decode
\newcommand{\RoneDecodeAboveMin}{0.6}  % figures/f7_r1.json:annotated.r1.frac_test_decode_above_unpaced_min
\newcommand{\RoneCalIdle}{9}  % results/declaration/heldout.json:results.6.cal_fp_by_workload.idle
\newcommand{\Digitiser}{PicoScope 5464E}  % paper/setup.json:digitiser.model
\newcommand{\DigBits}{16}  % paper/setup.json:digitiser.bits
\newcommand{\BurstRate}{100}  % paper/setup.json:aliasing.burst_rate_msps
\newcommand{\AliasTrainPct}{0.32}  % paper/setup.json:aliasing.train_folded_pct
\newcommand{\AliasIdleX}{13}  % paper/setup.json:aliasing.idle_folded_ratio
\newcommand{\BootBlock}{12}  % paper/setup.json:intervals.block_windows
\newcommand{\BootBlockS}{60}  % paper/setup.json:intervals.block_s
\newcommand{\NBoot}{2,000}  % paper/setup.json:intervals.n_boot
\newcommand{\EpBlock}{4}  % paper/setup.json:intervals.episode_block
\newcommand{\OnsetBlockMin}{10}  % paper/setup.json:intervals.onset_block_min
\newcommand{\TfourIdleADecode}{99.97}  % results/robustness/t4.json:misspecified_declaration.declared.h3.r1.false_alarms.idle.A
\newcommand{\NBootBound}{10,000}  % paper/setup.json:intervals.n_boot_bound
\newcommand{\ForestTrees}{500}  % paper/setup.json:forest_trees
\newcommand{\TauRoneLo}{0.545}  % results/heldout/thresholds.json:r1.stability.ci95.0
\newcommand{\TauRoneHi}{0.641}  % results/heldout/thresholds.json:r1.stability.ci95.1
\newcommand{\AlphaPct}{5}  % results/heldout/h2.json:alpha
\newcommand{\EthreeKOne}{4}  % results/heldout/e3_calibration_breakdown.json:breakdown_by_alpha.0.01.n_windows
\newcommand{\EthreeKFive}{18}  % results/heldout/e3_calibration_breakdown.json:breakdown_by_alpha.0.05.n_windows
\newcommand{\EthreeKTen}{38}  % results/heldout/e3_calibration_breakdown.json:breakdown_by_alpha.0.1.n_windows
\newcommand{\EthreeCalWindows}{340}  % results/heldout/e3_calibration_breakdown.json:calibration_windows
\newcommand{\EthreeHonestMin}{99.66}  % results/heldout/e3_calibration_breakdown.json:breakdown_honest_min
\newcommand{\DecodeRuns}{5}  % results/heldout/e3_calibration_breakdown.json:decode_summary.n_runs
\newcommand{\DecodeSessions}{4}  % results/heldout/e3_calibration_breakdown.json:decode_summary.n_sessions
\newcommand{\DecodeWindows}{850}  % results/heldout/e3_calibration_breakdown.json:decode_summary.n_windows
\newcommand{\DecodeFlagged}{5}  % results/heldout/e3_calibration_breakdown.json:decode_summary.n_flagged
\newcommand{\PrefillInFitFlagged}{1}  % results/heldout/e3_calibration_breakdown.json:dev_prefill_in_fit.n_flagged
\newcommand{\PrefillInFitN}{170}  % results/heldout/e3_calibration_breakdown.json:dev_prefill_in_fit.n
\newcommand{\RoneUnpacedMin}{0.658}  % results/heldout/e3_calibration_breakdown.json:unpaced_margin.run_min
\newcommand{\RoneMargin}{0.017}  % results/heldout/e3_calibration_breakdown.json:unpaced_margin.margin_over_tau_ci95_hi
\newcommand{\DeclBandLo}{0.39}  % results/heldout/e3_calibration_breakdown.json:score_band_min_median_max.full
\newcommand{\DeclBandHi}{0.68}  % results/heldout/e3_calibration_breakdown.json:score_band_min_median_max.full
\newcommand{\HidBandLo}{0.66}  % results/heldout/e3_calibration_breakdown.json:score_band_min_median_max.full
\newcommand{\HidBandHi}{0.71}  % results/heldout/e3_calibration_breakdown.json:score_band_min_median_max.full
\newcommand{\EfourKHonest}{50}  % results/heldout/e4_fit_contamination.json:grid.6.k
\newcommand{\EfourShareHonest}{7}  % results/heldout/e4_fit_contamination.json:grid.6.share_of_fit
\newcommand{\EfourFullCal}{0}  % results/heldout/e4_fit_contamination.json:grid.6.calibration_honest_by_run.panel-core-training-r4.median
\newcommand{\EfourLoraCal}{100}  % results/heldout/e4_fit_contamination.json:grid.6.calibration_honest_by_run.panel-core-lora-r4.median
\newcommand{\EfourKHalf}{500}  % results/heldout/e4_fit_contamination.json:smallest_k.0.5.k
\newcommand{\EfourKGone}{1000}  % results/heldout/e4_fit_contamination.json:smallest_k.0.05.k
\newcommand{\EfourShareGone}{60}  % results/heldout/e4_fit_contamination.json:smallest_k.0.05.share_of_fit
\newcommand{\EfourNDraws}{20}  % results/heldout/e4_fit_contamination.json:grid.6.n_draws
\newcommand{\EfourNDeclared}{680}  % results/heldout/e4_fit_contamination.json:n_declared_fit
\newcommand{\TfourPosCtrl}{0.17}  % results/heldout/e3_calibration_breakdown.json:decode_only_flag_rates.honest training.pooled
\newcommand{\RoneHonestFull}{100}  % results/heldout/e3_calibration_breakdown.json:curve_alpha_0.05.0.rates.honest training
\newcommand{\DecOnlyPrefillPooled}{0}  % results/heldout/e3_calibration_breakdown.json:decode_only_flag_rates.declared development serving-prefill.pooled
\newcommand{\EfourHalfDet}{35}  % results/heldout/e4_fit_contamination.json:grid.9.unpaced_detection.median
\newcommand{\EfourGoneDet}{0}  % results/heldout/e4_fit_contamination.json:grid.10.unpaced_detection.median
\newcommand{\SessSame}{11}  % results/heldout/session_check.json:results.iforest.pooled.S
\newcommand{\SessOther}{29}  % results/heldout/session_check.json:results.iforest.pooled.O
\newcommand{\SessPair}{22}  % results/heldout/session_check.json:results.iforest.per_run.panel-core-serving-decode-r1.other_run_fp_by_run.panel-core-serving-decode-r4
\newcommand{\SessRefWindows}{10}  % results/heldout/session_check.json:design.n_ref
\newcommand{\SessRuns}{3}  % results/heldout/session_check.json:decode_runs
\newcommand{\FeightBin}{0.25}  % results/monitor/traces.json:envelope.bin_s
\newcommand{\RobCoMin}{100}  % results/robustness/t4.json:alpha_sweep
\newcommand{\RobOnDecAlpha}{0}  % results/robustness/t4.json:alpha_sweep
\newcommand{\RobOnIdleAlpha}{4}  % results/robustness/t4.json:alpha_sweep
\newcommand{\RobOnDecPers}{2}  % results/robustness/t4.json:persistence
\newcommand{\RobOnIdlePers}{4}  % results/robustness/t4.json:persistence
\newcommand{\EfourDeclZero}{0.6}  % results/heldout/e4_fit_contamination.json:grid.0.declared_fit_flag_rate.median
\newcommand{\EfourDeclEnd}{0.0}  % results/heldout/e4_fit_contamination.json:grid.-1.declared_fit_flag_rate.median
\icmltitlerunning{Verifying Declared AI Workloads from GPU Supply Current}

\begin{document}

\twocolumn[
  \icmltitle{Verifying Declared AI Workloads from GPU Supply Current}

  \begin{icmlauthorlist}
    \icmlauthor{Tom Kimpson}{unimelb,mats}
  \end{icmlauthorlist}

  \icmlaffiliation{unimelb}{School of Mathematics and Statistics, The University of Melbourne, Parkville, VIC, Australia}
  \icmlaffiliation{mats}{Machine Alignment, Transparency \& Security (MATS)}

  \icmlcorrespondingauthor{Tom Kimpson}{tom.kimpson@unimelb.edu.au}

  % Keywords populate the PDF metadata; they are not shown in the document.
  \icmlkeywords{AI governance, compute verification, declared workloads, hidden training, side channels, current sensing, anomaly detection, one-class models}

  \vskip 0.3in
]

% Creates the first-column footnote with affiliations; required even if empty.
\printAffiliationsAndNotice{}

\begin{abstract}
Frontier AI governance agreements require compute verification techniques that allow an external observer to confirm that a hardware operator does not run prohibited workloads. Off-chip, analogue hardware signals, such as power draw, have been proposed as a potential information channel for compute verification. Such signals can be measured externally, without the operator's cooperation. We show that the supply current drawn by a GPU can be used to verify the work declared by the GPU operator, without requiring examples of prohibited workloads or evasion strategies. We record \CorpusHours{}~h of GPU current with a Rogowski coil under idle, training and inference workloads. We consider five evasion strategies an operator might use to disguise prohibited training. We show that a one-class model fitted to the current draw of permitted, declared work (idle and inference) flags at least \RoneMinGroup\% of windows of prohibited training, including training hidden by the five evasion strategies. As a continuous monitor across successive windows, it detects all \EpScored{} episodes of prohibited training hidden by time-sharing with declared inference, with a median delay of \EpDelayRone{}~s, and raises one alarm in \ExpDeclH{}~h of declared work. Incomplete or corrupt recordings of the declared work can undermine verification. An incomplete recording fails loudly; with idle missing, the monitor is in alarm for \TfourIdleADecode\% of declared idle. A corrupt recording can fail silently; if the operator runs training during the recording, the monitor can miss hidden training while raising no alarm on declared work. These results suggest that operator declarations together with external analogue sensing might offer a practical route to compute verification. 
\end{abstract}

\section{Introduction}\label{s:intro}

Agreements on AI compute will need methods to verify how GPUs are used \citep{shavit2023,baker2025}. The operator states the work that a GPU runs, its
\emph{declaration}. The verifier does not trust the declaration, and must check that
the GPU runs nothing else, such as hidden training. 

Off-chip analogue signals, such as power draw, can be measured externally by the verifier. There are two broad approaches to using such signals for compute verification \citep{scher2025}. In the \emph{default approach}, the verifier runs labelled examples of permitted and
prohibited work on its own chips, including prohibited work disguised by the evasion
strategies it anticipates, and fits a classifier to the resulting signals. The classifier
is then applied to the operator's chips. It must transfer from the chips it was fitted on
to the chips it monitors, and it is fitted only to the evasions the verifier anticipated. In the alternative approach, which we term the \emph{signature check}, the verifier, with temporary access to the operator's hardware, records signatures of permitted workloads. Later, the verifier monitors the chips' activity and flags any signals that depart from the recorded signatures. This approach needs no examples of prohibited work and no classifier that transfers
between chips. It is also harder to evade, because the signals must look specifically like the declared
work, rather than merely unlike training. \citet{baker2025} propose a hybrid, in which signatures
of permitted work are recorded on the verifier's own hardware, and so must transfer to the
operator's chips.

Existing empirical studies of verification to identify hidden training have focused on the default approach.
\citet{gargiulo2026}
measure the current drawn by a GPU. A random forest fitted to current traces identifies
training from inference and non-AI work with 97\% accuracy on model families it has
not seen. Similarly, \citet{rahman2026} reach 98\% from sliding windows of on-chip telemetry
across nine GPU types. Both studies test their classifiers against operators who try to
hide training. Detection stays high against most evasion strategies, but falls against some that
the classifiers were not fitted to: to 43\% for white-box diluted low-rank adaptation in
\citet{rahman2026}, and to 48--88\% for diluted low-rank adaptation in
\citet{gargiulo2026}, though a second test there, which flags each optimiser update,
restores detection to at least 98\%.

We found no empirical study of a signature check built on off-chip signals. Power signatures do
identify workloads on high-performance computing systems~\citep{copos2020,kohler2021,karimi2024},
but those studies assume an operator who does not try to hide work. There are several uncertainties in the signature check \citep{scher2025}: the verifier must choose which chip activity to measure, and how stable signatures stay over time; the verifier must decide how often to investigate an alarm, how finely to define workloads, and whether the chip activity of each workload is specific enough to tell it apart.

In this work we test the signature check on one NVIDIA \GPU{}. A current probe is clamped to the GPU's auxiliary supply and records \CorpusHours{}~h of current
(\cref{s:setting}). We record declared work
(serving decode, serving prefill and idle), honest training, and several evasion
strategies, in particular \emph{co-location}: training runs as a second process beside
the declared inference and uses the capacity that inference leaves spare. We fit
classifiers to labelled examples of training to stand for the default approach
(\cref{s:missed}). Our signature
check is a one-class model fitted only to recordings of the declared work, with its threshold set
on separate recordings of that work, so it needs no examples of training or evasion. Together these recordings form its
\emph{reference}. We first test the check with a complete reference (\cref{s:works}).
The signature check is only as good
as its reference. A declared workload can be missing from the reference, or else an operator can try to corrupt the reference, for example by running training while it is recorded. We therefore
test the signature check with an incomplete reference and with the reference corrupted by training (\cref{s:monitoring}). We also run the signature check continuously over a held-out test day, as a monitor, and ask whether a verifier can tell from its alarms that the reference is incomplete or corrupt. We measure four of the uncertainties listed above(\cref{s:declaration}). 
We close with the lessons for verifiers (\cref{s:discussion}).

\section{Methods}\label{s:setting}

\subsection{Measurement}
Every AI workload is executed by a single \GPU{} hosted by \citet{amodo2026}. A Rogowski current probe (\Probe{},
\ProbeSens{}~mV/A, \ProbeLo{}~Hz to \ProbeHi{}~MHz) is attached to the GPU's auxiliary supply. A \DigBits-bit \Digitiser{} samples the probe at \Fs{}~S/s, with no anti-alias filter \footnote{Without an anti-alias filter, current above the Nyquist frequency (about 100 kHz) aliases into the recorded band. Under training this adds at most\AliasTrainPct \% to the in-band variance, but under idle it may exceed the in-band variance by up to \AliasIdleX{} times (\cref{a:impl}).} We note that the probe is AC-coupled, with no voltage channel, and so measures current rather than power. 

\subsection{Threat model}\label{s:threat}

The verifier receives the probe's trace and the operator's declaration of work. The verifier also holds pre-recorded current traces (the reference) of the hardware under permitted workloads. The verifier owns and controls the current probe. The operator controls the software on the GPU and cannot physically touch or interfere with the probe. The operator may run any workload, including the adversarial strategies of \cref{s:corpus}, to hide training. We do not test an adaptive operator that tunes its evasion strategies against the verifier's detector. We do test the edge case of an operator that corrupts the reference by e.g. running training while the verifier records the declared work (\cref{s:monitoring})

\subsection{Corpus}\label{s:corpus}
We use \NCaptures{} workload runs (\CorpusHours{}~h of current traces).

Earlier studies use shorter recordings; for example \citet{gargiulo2026} record \GkTraceS{}-s traces
at about \GkRateApprox{}~MS/s, one workload per trace, while \citet{rahman2026} use NVML
windows of \RtWinLo{} to \RtWinHi{}~s. In practice a verifier may observe a GPU for hours and a single evasion may start and stop inside a recording of declared work, so only continuous recordings let us test detectors as monitors
(\cref{s:works}). Every AI workload uses a decoder-only transformer, written in PyTorch, with \ModelLayers{} layers, width
\ModelWidth, and \ModelParams{}M parameters. The data is open-sourced at \citet{kimpson2026corpus}. The runs fall into the following workload groups:
\begin{itemize}
  \item \emph{honest training}: full fine-tuning and LoRA. Some full fine-tuning runs add
        checkpoint evaluation or a copying control, or discard every update;
  \item \emph{time-sharing}: training and inference alternate in slots;
  \item \emph{co-location}: training and inference run as two resident processes,
        and the driver time-slices them. Each run sets a target training share
        $\rho \in \{\RhoLo, \RhoMid, \RhoHi\}$, one run each. At $\rho = \RhoLo$
        and $\RhoHi$ one process pauses between steps to hold the share; we call
        these runs \emph{throttled}. At
        $\rho = \RhoMid$ neither process pauses, and the driver alone sets the share.
        We call this run \emph{unthrottled};
  \item \emph{jitter}, \emph{padding} and \emph{power cap}: training with
        perturbed step timing or a lowered power limit;
  \item \emph{declared work}, the workloads of the declaration: serving decode
        (generating tokens), serving prefill (reading prompts) and idle;
  \item \emph{non-AI work}: dense GEMM, memory-bound kernels and periodic decoys.
\end{itemize}
Time-sharing, co-location, jitter, padding and power cap are the \emph{evasion groups}.

The runs are split into three sets, each recorded on different days: $\mathcal{D}_{\mathrm{dev}}$ for development, $\mathcal{D}_{\mathrm{test}}$ for testing, and $\mathcal{D}_{\mathrm{cal}}$ for calibration. $\mathcal{D}{\mathrm{test}}$ is one day, the \emph{test day}; $\mathcal{D}{\mathrm{cal}}$ is the day after it, and $\mathcal{D}{\mathrm{dev}}$ every day before it. \Cref{t:corpus} summarises the runs and traces, factorised by workload group and set.

\subsection{Windows and features}
A window $x$ is \Win{}~s of trace. The hop between consecutive windows is \HopSnap{}~s
for window-level results and \HopMon{}~s for the monitor. We define a window as
\emph{training} iff it holds at least one
complete optimiser step, following \citet{gargiulo2026}. Windows with only part
of a step are left out of detection rates. Each window gives \NFeat{} amplitude-free
features of the raw spectrum and of the envelope $e(t)$, the RMS of the raw trace over
\EnvFrame{}~ms frames. They are adapted from \citet{gargiulo2026}, who chose them to
separate training from inference.

\subsection{Detectors}\label{s:detectors}
Each detector assigns a score $s(x)$ to a window $x$. Our detectors follow the two
approaches of \citet{scher2025}.

The \emph{signature check} $s_R$ is an isolation
forest~\citep{liu2008}. It is fitted to the declared windows of
$\mathcal{D}_{\mathrm{dev}}$ only: decode, prefill and idle. Its score is the negated isolation score, so a high $s_R$ means ``unlike the
declaration''. It never sees training, evasion or non-AI work.

The \emph{classifier} is the two-stage random forest of \citet{gargiulo2026}, with
\ForestTrees{} trees per stage (settings in \cref{a:impl}). The first stage separates AI from non-AI work. The AI-classified work is then passed on to the second stage, which separates training from inference. The classifier stands for the default approach, because it learns from labelled examples of training. Because it is fitted on the GPU it monitors, it is the most favourable case of that approach (\cref{s:missed}). The classifier's score is
\begin{equation}
  s_C(x) = \begin{cases}
    \hat P(\text{training} \mid x, \text{AI}) & \text{stage one calls } x \text{ AI},\\
    0 & \text{otherwise}.
  \end{cases}
\end{equation}
We fit the classifier in three ways:
\begin{itemize}
  \item $s_C^{\mathrm{dev}}$ is fitted to $\mathcal{D}_{\mathrm{dev}}$ only. So that each
        evasion is unseen when scored, it is one fit per evasion group, with that
        group left out, and each group is scored by the fit that never saw it.
        ``The classifier'' means $s_C^{\mathrm{dev}}$;
  \item the \emph{no-evasion fit} is $s_C^{\mathrm{dev}}$ with every evasion group
        removed;
  \item the \emph{replica} is fitted to \NRepTraces{} traces
        recorded following the design of \citet{gargiulo2026}, where it reaches
        macro-$F_1 = \SoneFone$ under leave-one-group-out.
\end{itemize}
$s_C^{\mathrm{dev}}$ and the no-evasion fit are fitted to our corpus, so the declared
work is among their examples of work that is not training. The replica is fitted to
none of the \NCaptures{} runs of \cref{s:corpus}: its traces, each of a single workload, were recorded separately on the
same GPU and probe.

\subsection{Thresholds}
A detector flags $x$ iff $s(x) > \tau_\alpha$, with
\begin{equation}\label{e:tau}
  \tau_\alpha = Q_{1-\alpha}\bigl\{ s(x) : x \in \mathcal{D}_{\mathrm{cal}}^{\mathrm{decl}} \bigr\}, \quad \alpha = \Alpha.
\end{equation}
$Q_{1-\alpha}$ is the empirical quantile, with linear interpolation, and
$\mathcal{D}_{\mathrm{cal}}^{\mathrm{decl}}$ is the declared windows of
$\mathcal{D}_{\mathrm{cal}}$, one decode run and one idle run of equal length. $s_R$ and every
classifier fit use the same $\tau_\alpha$ rule. For a set $W$ of
training windows, detection is
$d = |W|^{-1} \sum_{x \in W} \mathbb{I}[s(x) > \tau_\alpha]$. The false-positive rate
is $d$ with $W$ the declared windows of $\mathcal{D}_{\mathrm{test}}$. Every fit of
$s_C^{\mathrm{dev}}$ has the same $\tau_\alpha$. Its window-level false-positive rates
come from the fit without co-location, which has the highest decode false-positive
rate of the five.

For $s_R$, $\tau_\alpha$ is our choice of rule for what \citet{scher2025} call
``sufficiently close to the baseline''. They
specify one baseline, whereas we use two separate recordings of declared work: the fit
($\mathcal{D}_{\mathrm{dev}}$) and the threshold set
($\mathcal{D}_{\mathrm{cal}}^{\mathrm{decl}}$). Together they form the \emph{reference} of
$s_R$. We tested no other rule. We varied only $\alpha$ (\cref{s:works}).

\subsection{Monitor}\label{s:monitor}
We run each detector over the test day as a monitor (\cref{f:f8}). This represents a verifier that
observes the current trace continuously. The verifier does not raise an alarm on a
single flagged window. Instead, an alarm starts when 3 of 5 consecutive windows are
flagged, and ends after \Rearm{} unflagged windows in a row;
\cref{s:works} varies this rule. The classifier's monitor uses the fit without
time-sharing, because the test day's embedded training episodes are time-shared
(\cref{s:works}). We describe false
alarms by two numbers: $F$, the alarm onsets per monitored hour, and $A$, the fraction
of time in alarm.

\section{Results}\label{s:results}

\subsection{The check with a complete reference}\label{s:works}

\begin{figure*}[t]
  \centering
  \includegraphics[width=\linewidth]{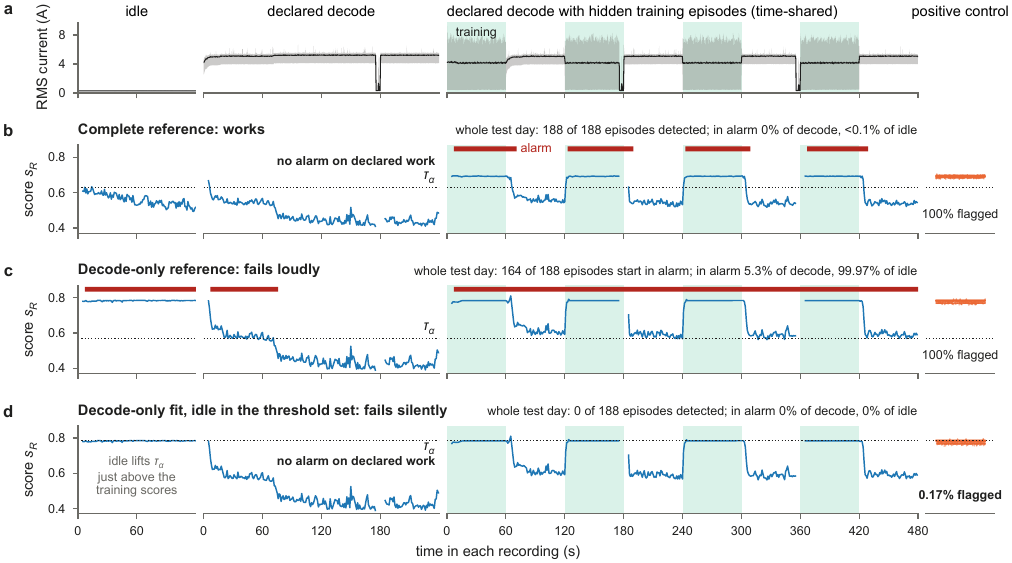}
  \caption{Three configurations of the signature check on the same test-day recordings.
    (a) RMS current in three separate recordings: idle, declared decode, and decode with
    time-shared training episodes (shaded). Line: mean of $e(t)$ per \FeightBin~s; band:
    its 5--95\% range. The dips are unscored seconds. (b)--(d) The score $s_R$, the
    threshold $\tau_\alpha$ (dotted) and the alarm (red bar). (b) The fit holds decode,
    prefill and idle; the threshold set holds decode and idle. (c) Both hold decode only.
    (d) The fit holds decode only; the threshold set holds decode and idle. Right column:
    $s_R$ of honest-training windows from the development data, a positive control.}
  \label{f:f8}
\end{figure*}

The signature check needs no examples of hidden training. With the complete reference, whose fit
holds decode, prefill and idle, $s_R$ at $\tau_\alpha = \TauRone$ detects
$\Ztwo\%$ of co-location windows (development data, one run per training share), and at
least \RoneMinGroup\% of every other evasion group, although it is fitted to
none of them. Its test false-positive rates are \FprRoneDecode\% on decode
and \FprRoneIdle\% on idle.

We also run $s_R$ over the test day as a monitor (\cref{s:monitor}; \cref{f:f8}b).
That day embeds \EpTotal{} time-shared training episodes of \EpMinDur{}~s to
\EpMaxDur{}~s inside declared inference. An episode is detected iff an alarm is active
at some time $t \in [t_{\mathrm{s}},\, t_{\mathrm{e}} + \Grace~\mathrm{s}]$, where its
first step starts at $t_{\mathrm{s}}$ and it ends at $t_{\mathrm{e}}$. An episode with
no alarm is missed, even if it is too short to hold a complete step. We leave out
\EpTruncated{} episodes that the recording cuts short. Both $s_R$ and the classifier, $s_C^{\mathrm{dev}}$,
detect all of the other \EpScored. For $s_R$ the median delay from
first step to alarm onset is $D = \EpDelayRone$~s, and even the shortest episodes
(\LminRone~s) have a within-run lower bound on detection above the target level
of \LminLevel. $s_R$ raises no alarm on test-day decode (\cref{t:t2}). These episodes
test time-sharing only; no co-located episode was recorded.

These results do not hinge on $\alpha$ or on the alarm rule. For $\alpha$ from 1\% to 10\%,
$s_R$ detects \RobCoMin\% of co-location windows and all \EpScored{} episodes, with
\RobOnDecAlpha{} false-alarm onsets on test-day decode and at most \RobOnIdleAlpha{} on idle. Under
persistence rules from 2-of-3 to 5-of-5 it still detects all \EpScored{} episodes, with at
most \RobOnDecPers{} onsets on decode and \RobOnIdlePers{} on idle.

\subsection{The check with an incomplete or corrupt reference}\label{s:monitoring}

\begin{figure}[t]
  \centering
  \includegraphics[width=\columnwidth]{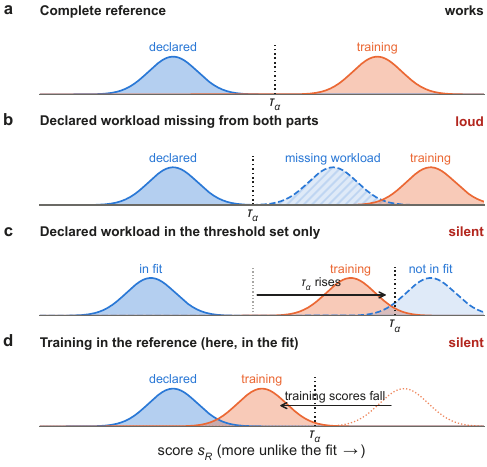}
  \caption{The complete reference (a) and the three changes to it (b--d). Each curve is the distribution of
    the score $s_R$ for one kind of work: declared work (blue) or hidden training
    (orange). The monitor flags every window to the right of the threshold $\tau_\alpha$
    (dotted). (a) The fit and the threshold set hold every declared workload. Training is
    flagged, and declared work is not. (b) One declared workload (dashed) is in neither. It
    scores high and is flagged: a loud failure. (c) The same workload is in the threshold set
    but not in the fit. It still scores high, so $\tau_\alpha$ rises to meet it and
    passes training. Training is no longer flagged: a silent failure. Training in the
    threshold set acts the same way. (d) Training is in the fit. Its scores fall below
    $\tau_\alpha$, so the monitor misses it: also a silent failure. \Cref{t:t7,f:f9} hold the measured cases.}
  \label{f:f10}
\end{figure}

\begin{table*}[tp]
  \centering
  \small
  \resizebox{\textwidth}{!}{% generated by scripts/table_t7.py from (repo_commit of each source):
% h2 41b1fed, r1_scores 53f9463, episodes c574166, false_alarms 1fbe999, heldout dcd87a8, t4 1658809, manifest 018c3a2
% cell sources:
%   full: fit r1_scores.fit_runs.r1; tau set r1_scores.calibration_runs; co-location h2.headline.Z_r1.detection; episodes episodes.detectors.r1.counts; onsets false_alarms.backgrounds.{declared inference,idle}.r1.onsets
%   prefill held out: fit heldout.fit_runs minus panel-serving-prefill; tau set heldout.calibration_runs; flagged heldout.results[family=iforest,variant=R1,alpha=0.05].check4.per_run.panel-serving-prefill.fp_rate; co-location, episodes, onsets: none
%   decode only, decode: fit t4.misspecified_declaration.fit_runs; tau set t4.misspecified_declaration.calibration_runs.declared; co-location t4.misspecified_declaration.declared.h2.r1.co-location.pooled; episodes t4.misspecified_declaration.declared.h3.r1.episodes.counts; flagged t4.misspecified_declaration.declared.dev_stress_fpr; onsets t4.misspecified_declaration.declared.h3.r1.false_alarms.{declared inference,idle}.onsets
%   decode only, decode + idle: fit t4.misspecified_declaration.fit_runs; tau set t4.misspecified_declaration.calibration_runs.pooled; co-location t4.misspecified_declaration.pooled.h2.r1.co-location.pooled; episodes t4.misspecified_declaration.pooled.h3.r1.episodes.counts; flagged t4.misspecified_declaration.pooled.dev_stress_fpr; onsets t4.misspecified_declaration.pooled.h3.r1.false_alarms.{declared inference,idle}.onsets
\begin{tabular}{lllrrlrl}
\toprule
reference & unlike the fit & in $\tau$ set & co-location (\%) & episodes & unlike-fit work flagged (\%) & onsets dec.\ / idle & failure \\
\midrule
complete reference & -- & -- & 100.0 & 0 of 188 missed & -- & 0 / 1 & none \\
prefill held out of the fit & prefill & none & -- & -- & prefill 100.0 & -- & loud \\
decode-only fit, decode threshold set & prefill, idle & none & 100.0 & 164 of 188 start in alarm & prefill 100.0, idle 100.0 & 2 / 4 & loud \\
decode-only fit, decode + idle threshold set & prefill, idle & idle & 0.0 & 188 of 188 missed & prefill 0.0, idle -- & 0 / 0 & silent \\
\bottomrule
\end{tabular}
}
  \caption{The signature check under four references, one per row. ``Unlike the
    fit'': declared work that is not in the fit. ``In $\tau$ set'':
    the part of it that is in the threshold set. Co-location: windows detected at
    $\tau_\alpha$. Episodes: the \EpScored{} scored episodes of the test day; ``start in
    alarm'' means that an alarm is already on when the episode starts (\cref{a:impl}).
    Flagged: windows of the unlike-fit work above $\tau_\alpha$. Onsets: alarm onsets on
    test-day decode and idle.}
  \label{t:t7}
\end{table*}

\begin{figure}[t]
  \centering
  \includegraphics[width=\columnwidth]{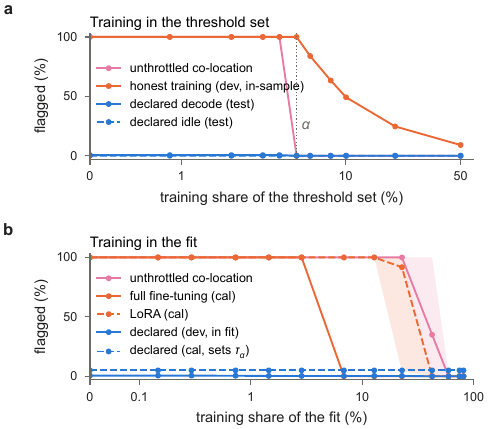}
  \caption{Training in the reference blinds $s_R$. (a) Honest-training windows added to the threshold set, with the
    fit unchanged and $\alpha = \AlphaPct\%$. The unthrottled run drops to zero near a
    training share of $\alpha$. Honest training is in-sample, since the added windows
    come from its pool. Test decode and idle are flagged less as $\tau_\alpha$ rises.
    (b) Honest-training windows added to the fit, with the threshold set unchanged:
    median over \EfourNDraws{} draws, band from the lowest to the highest draw. Full
    fine-tuning and LoRA are calibration-day runs, outside every fit. The development
    declared windows are in every fit (in-sample). The calibration declared windows set
    $\tau_\alpha$, so their rate stays near $\alpha$ by construction.}
  \label{f:f9}
\end{figure}

The reference is where the signature check can be attacked by an adversarial operator. It is recorded on the
operator's chips, so an operator can try to corrupt it, and even a reference recorded in good
faith can erroneously leave out a declared workload.\footnote{For example, the operator
declares decode, prefill and idle, but the inspectors record only decode. Nobody cheats,
yet the reference is incomplete.} The reference has two parts: the fit, and the threshold set
that fixes $\tau_\alpha$. In \cref{s:works} the fit holds every declared workload, and
the threshold set holds decode and idle.

Here we change the reference in three ways (\cref{f:f10}b--d). Each
change makes the check fail, and each failure is one of two kinds. A \emph{loud} failure
raises alarms on declared work, so the verifier sees that the reference is wrong. A
\emph{silent} failure misses hidden training and raises no alarm on declared work, so
the verifier sees nothing.

\textbf{A declared workload missing from both parts: loud} (\cref{f:f10}b,
\cref{t:t7}). Work that is in neither the fit nor the threshold set raises alarms.
Prefill held out of both is one case (\cref{s:declaration}). A decode-only reference,
with decode alone in the threshold set, is another (\cref{f:f8}c). It detects
\TfourMisCoDeclared\% of co-location, but it is also in alarm for \TfourIdleADecode\% of
test-day idle. A monitor that is almost always in alarm on declared idle is useless, but
it fails loudly. Its alarms show the verifier that it is failing.

\textbf{A declared workload in the threshold set only: silent} (\cref{f:f10}c,
\cref{t:t7}). Keep the decode-only fit, and add idle to its threshold set. The threshold set
now holds decode and idle, which is our default. Idle is now in the threshold set
but not in the fit. It scores high
and lifts $\tau_\alpha$ to \TauDecOnlyPooled{} (\cref{f:f8}d; \cref{f:f7} shows the
scores). The monitor then detects \TfourMisCoPooled\% of co-location and misses all
\EpScored{} episodes. It raises no alarm on test-day decode, just as the complete reference
does. We see that this monitor then fails silently; it does not alert on hidden training, and the verifier has no way to know that it is failing. Prefill is also missing from this fit. Accordingly, with decode alone in the threshold set every prefill window is flagged (\cref{t:t7}). However $\tau_\alpha$ now sits above the prefill scores, and so \DecOnlyPrefillPooled\% of prefill windows are flagged. That is, a missing workload raises alarms only if nothing else
has raised the threshold.

A verifier can expose the silent monitor with a positive control. It gives the monitor
work that it knows to be training, and checks that the monitor flags it. The working
monitor flags \RoneHonestFull\% of development honest training. The silent monitor flags
\TfourPosCtrl\% (\cref{f:f8}, right column).

\textbf{Training in the reference: silent} (\cref{f:f10}c--d, \cref{f:f9}). An operator can
also corrupt the reference, by running training while it is recorded. We test this by
adding windows of training that runs alone on the GPU (honest training,
\cref{s:corpus}) to each part of the reference in turn (\cref{a:reference}). In both cases the monitor comes to miss hidden training, and it
raises no alarm on declared work.

\emph{Training in the threshold set raises $\tau_\alpha$} (\cref{f:f10}c). It takes little training. $\tau_\alpha$ is the $(1-\alpha)$ quantile of the
threshold set (\cref{e:tau}), so once training makes up a fraction $\alpha$ of that set,
$\tau_\alpha$ sits among the training scores. At
$\alpha = \AlphaPct\%$, \EthreeKFive{} windows added to the \EthreeCalWindows{}
calibration windows take detection of the unthrottled run to zero (\cref{f:f9}a). The
higher $\tau_\alpha$ flags declared work less, not more, so the failure is silent. A
positive control of honest training does not expose it, because at that share honest
training is still flagged at $\ge \EthreeHonestMin\%$.

\emph{Training in the fit lowers the scores of training} (\cref{f:f10}d). The fit learns
training as if it were declared work. How much training this takes depends on its kind.
With training at \EfourShareHonest\% of the fit, a full fine-tuning run outside the fit
is flagged at \EfourFullCal\%. The unthrottled run holds longer. Its median detection
reaches \EfourGoneDet\% at a share of \EfourShareGone\% (\cref{f:f9}b). The declared
windows of the fit are flagged at \EfourDeclZero\% with a clean fit, and less as training
is added, so again no alarm shows the failure. This last result is in-sample, because we did not
score held-out declared work under these fits.

\subsection{The open questions}\label{s:declaration}

A verifier who adopts the signature check must still settle four questions
raised by \citet{scher2025} (\cref{s:intro}). Does the signature stay stable
over time? How often will the check raise an alarm? How finely must workloads
be defined? And is the signature specific enough to separate declared work from
hidden work? We answer each in turn from the corpus of \cref{s:corpus}.

\paragraph{Stability over time.} The signature of decode stayed stable over the five days of recording. We recorded
decode in \DecodeRuns{} runs over \DecodeSessions{} sessions, each on a different day.
Three runs are in the reference of $s_R$, two in its fit and one in its threshold set.
Across all \DecodeRuns{} runs, $s_R$ flags
\DecodeFlagged{} of \DecodeWindows{} windows at $\tau_\alpha$, and each flagged window
is the first window of its run. On the two test-day runs, which are outside the reference, its false-positive rate is
\FprRoneDecode\%. Idle was less stable. $s_R$ flags \RoneCalIdle\% of calibration
idle, but only \FprRoneIdle\% of test idle, recorded one day earlier. Because
calibration idle scores above decode, it sets $\tau_\alpha$, and the false-positive rate on test
decode stays well below $\alpha = \AlphaPct\%$.

A reference recorded on a single day does not carry over to other days. In an exploratory check, we
fitted a reference to \SessRefWindows{} windows spread over about four minutes of one
decode run, and repeated this for each of \SessRuns{} runs, one per day. Averaged over
the three runs, this reference flags \SessOther\% of decode from the other days,
against \SessSame\% of the rest of its own run. The two runs recorded to one protocol,
two days apart, give \SessPair\% each way. This reference is not directly comparable to the reference of $s_R$ because it is shorter and flags by a leave-one-out rule rather than $\tau_\alpha$. Nevertheless, this check suggests that the reference should be recorded over several sessions, as the reference of $s_R$ is.

\paragraph{Alarm frequency.} The alarm rate sets how often inspectors must investigate.
On the test day, declared inference is decode only, and the
monitors (\cref{s:monitor}) watch $E = \ExpDecodeH$~h of it (\cref{t:t2}):
\begin{itemize}
  \item $s_R$ raises no alarm: $F = \FRoneDecode$~h$^{-1}$ (95\% upper bound
        \FupRoneDecode~h$^{-1}$), $A = \ARoneDecode$;
  \item the classifier, $s_C^{\mathrm{dev}}$, raises $F = \FClfDecode$~h$^{-1}$ (upper bound
        \FupClfDecode~h$^{-1}$), $A = \AClfDecode$.
\end{itemize}
This exposure is short, so these rates bound the alarm rate only loosely. On idle
($E = \ExpIdleH$~h), both monitors stay near zero.

\begin{table}[tp]
  \centering
  \footnotesize
  % generated by scripts/table_t2.py from results/monitor/false_alarms.json (repo_commit 1fbe999)
% 3-of-5 monitor on 1 s-hop test windows; upper bound: exact Poisson model;
% A interval: approximate onset-block bootstrap (under-covers)
\setlength{\tabcolsep}{3pt}
\begin{tabular}{@{}lrrrr@{}}
\toprule
Detector & onsets & $F$ (/h) & upper (/h) & $A$ (\%) \\
\midrule
\multicolumn{5}{@{}l}{\emph{declared decode}, $E = 0.475$ h} \\
\quad $s_R$ & 0 & 0.0 & 6.3 & 0.0 [0.0, 0.0] \\
\quad $s_C^{\mathrm{dev}}$ & 3 & 6.3 & 16.3 & 1.0 [0.0, 2.0] \\
\midrule
\multicolumn{5}{@{}l}{\emph{idle}, $E = 8.350$ h} \\
\quad $s_R$ & 1 & 0.1 & 0.6 & 0.0 [0.0, 0.1] \\
\quad $s_C^{\mathrm{dev}}$ & 0 & 0.0 & 0.4 & 0.0 [0.0, 0.0] \\
\midrule
\multicolumn{5}{@{}l}{\emph{non-AI busy}, $E = 0.950$ h} \\
\quad $s_R$ & 4 & 4.2 & 9.6 & 99.8 [99.6, 99.9] \\
\quad $s_C^{\mathrm{dev}}$ & 0 & 0.0 & 3.2 & 0.0 [0.0, 0.0] \\
\bottomrule
\end{tabular}

  \caption{False alarms on the test day, per background, with exposure $E$.
    ``upper'': the 95\% upper bound on $F$, which assumes Poisson onsets. Overlapping
    windows make onsets correlated, so this bound is approximate. The intervals on $A$
    come from a block bootstrap (\cref{a:impl}). The test day holds no declared prefill. Non-AI work is undeclared, so $s_R$ is correct to alarm
    on it.}
  \label{t:t2}
\end{table}

\paragraph{Granularity.} The reference must hold each declared workload. When declared prefill
is left out of the reference of $s_R$, $s_R$ flags \RonePrefillHoldout\% of it (one
development run). This is expected, because prefill is limited by compute and decode
by memory bandwidth, so the two load the GPU differently. With prefill
in the fit, $s_R$ flags \PrefillInFitFlagged{} of \PrefillInFitN{} prefill windows.
This rate is in-sample, because neither the calibration day nor the test day holds
prefill. One detector, fitted to decode, prefill and idle together, can therefore
accept all three. The verifier must record every declared workload, but does not
need a separate detector for each. A workload that is in the
threshold set but not in the fit instead raises the threshold (\cref{s:monitoring}).

\paragraph{Specificity.} The signature is specific, but by a thin margin. Under the complete reference, declared
windows score from \DeclBandLo{} to \DeclBandHi{} on $s_R$, and co-location, honest
training and non-AI windows from \HidBandLo{} to \HidBandHi{} (\cref{f:f7}). The two
bands overlap only at the first window of each decode run. The hidden run that comes
closest to declared work is the unthrottled co-location run. Its lowest score,
\RoneUnpacedMin, is \RoneMargin{} above \TauRoneHi, the upper end of the 95\% interval
of $\tau_\alpha$ under a bootstrap of the two calibration runs. With two runs, that end
is the threshold that calibration idle alone would set. The signature check also
does not fully separate this run from declared decode. Its AUC is \RoneAucDecode, and
\RoneDecodeAboveMin\% of test decode windows score above the run's lowest score.

\subsection{Against the default approach}\label{s:missed}

A classifier, like the signature check, works only if its fit covers the declared work.
\Cref{t:t6} sets $s_R$ beside the three classifier fits. It lists what each detector
learns from, its detection of the unthrottled co-location run at $\tau_\alpha$, and its
false-positive rate on test decode.

\begin{table*}[!tp]
  \centering
  \small
  % generated by scripts/table_t6.py from (repo_commit of each source):
% operating_point.json 1d2e0a2, h2.json 41b1fed, r1_scores.json 53f9463,
% windows/manifest.json 018c3a2, replication/t3.json (amodo 73ffebe)
\begin{tabular}{lccccrrl}
\toprule
 & \multicolumn{4}{c}{learns from} & & & \\
\cmidrule(lr){2-5}
detector & declared & honest training & evasions & non-AI & $d_{\mathrm{unthrottled}}(\tau_\alpha)$ & FPR decode & alarm means \\
\midrule
replica & pub. & pub. & -- & pub. & 0.6\% & 0.0\% & training \\
no-evasion fit & \checkmark & \checkmark & -- & \checkmark & 100.0\% & 5.3\% & training \\
$s_C^{\mathrm{dev}}$ & \checkmark & \checkmark & \checkmark & \checkmark & 100.0\% & 1.5\% & training \\
$s_R$ & \checkmark & -- & -- & -- & 100.0\% & 0.6\% & undeclared \\
\bottomrule
\end{tabular}

  \caption{The data each detector learns from, its detection of the unthrottled run at
    $\tau_\alpha$, and its false-positive rate on test decode. ``pub.'': traces recorded
    on the same GPU and probe to the published design, not among the \NCaptures{}
    runs of \cref{s:corpus}. An alarm of $s_R$ means ``undeclared''; an alarm of a
    classifier means ``training''. The rows for the replica and the no-evasion fit are
    exploratory.}
  \label{t:t6}
\end{table*}

At its $\tau_\alpha$, the classifier, $s_C^{\mathrm{dev}}$, does nearly as well as $s_R$.
It detects $\Xprime\%$ of co-location, and every window of every other group except
jitter, where its weakest run is detected at \ClfJitterWorst\%. It stands for the
default approach because it learns from examples of training. But it is fitted on the
monitored GPU, declared work included, as the signature check is, so it needs no
transfer between chips. It is therefore the most favourable case of the default
approach. Even without evasion examples, as in the no-evasion fit, a classifier
needs examples of honest training, whereas $s_R$ needs only recordings of the declared
work.

The threshold of $s_C^{\mathrm{dev}}$ is also fragile. The classifier scores
\FracCalZero\% of calibration declared windows exactly zero, so
$\tau_\alpha = \TauClf$. It then flags any window that stage one calls AI and that even
one of its \ForestTrees{} stage-two trees votes as training. Small changes in declared
work could therefore shift its flags.

A classifier gives no guarantee on a declared workload that its fit has not seen. The
replica, fitted to traces recorded to the published design, detects only
\XzeroPrime\% of the unthrottled run at its $\tau_\alpha$, and flags \RepPrefill\% of
development prefill. $s_C^{\mathrm{dev}}$, which has development prefill in its fit,
flags \ClfDevPrefill\% of it. \citet{gargiulo2026} publish two forests fitted to their
own corpus: a genuine forest, fitted to their training, inference and non-AI traces,
and a hardened forest, which adds their adversarial traces. The two disagree on our
windows (\cref{t:t5}). At their default threshold of \DefaultDecision, the hardened
forest catches all three co-location runs, but it flags \GkHardPrefill\% of declared
prefill windows as training. The genuine forest flags only \GkGenPrefill\% of prefill,
but it misses the unthrottled run ($\GkGenMiss\%$). Nothing in either fit tells the
verifier which outcome to expect. The replica's traces differ from ours in workload
design and session, so its results do not rank the signature check against the default
approach.

\section{Discussion}\label{s:discussion}

The signature check worked when its reference held every declared workload. With no
example of hidden training in its fit, $s_R$ flagged every evasion group, and it
raised no alarm on test-day decode (\cref{s:works}). When the reference missed a declared
workload, the check failed in one of two ways (\cref{s:monitoring}). A workload missing
from both parts of the reference kept the monitor in alarm almost continuously while it
ran. This failure was loud, so the verifier could see that something was wrong. A workload in
the threshold set but not in the fit raised $\tau_\alpha$ and blinded the monitor. That
monitor raised no more alarms on declared work than the working one, so the failure was
silent. Such a workload could also mask a loud failure. Training in the reference also blinded the
monitor. So a quiet monitor is no evidence of a working one. The signature check needs no examples of hidden
training in its fit, but the verifier still needs them as positive controls, to test that
the monitor works.

\paragraph{Acquiring a reference and auditing the monitor.} The signature check is only as good as
its reference. The reference is recorded on the monitored chip, where the operator
controls the work. The verifier must therefore control how it is recorded.
\begin{itemize}
  \item \emph{Witness it.} Our reference recordings are our own honest runs. A real
        verifier needs recordings during which the operator cannot train. Both parts of
        the reference are targets. Training at a share of about $\alpha$ in the threshold
        set takes detection of the unthrottled run to zero, and full fine-tuning at
        \EfourShareHonest\% of the fit hides full fine-tuning (\cref{s:monitoring}).
  \item \emph{Span sessions.} A reference from one session flags \SessOther\% of decode
        from the other days (\cref{s:declaration}).
  \item \emph{Test with held-out declared workloads, not only held-out attacks.} A
        declared workload missing from the reference raises alarms
        (\cref{s:declaration}), unless a workload in the threshold set but not in the
        fit masks it (\cref{s:monitoring}).
  \item \emph{Audit the monitor with positive controls that sit close to the
        reference.} The silent monitor flags \TfourPosCtrl\% of honest training, so one
        control run exposes it. But a monitor that flags a control can still miss other hidden work. At the share
        of training in the threshold set that hides the unthrottled run, honest training
        is still flagged (\cref{s:monitoring}). Honest training sits far from the
        reference, so it cannot expose a raised threshold. A control must sit as close
        to the reference as the hidden work it represents.
\end{itemize}
It is hard to detect anomalous behaviour when permitted behaviour is open-ended \citep{sommer2010}. An agreement can close permitted behaviour off by requiring the operator to declare every workload, so that
anything else is an anomaly by rule. Declarations also help verification more broadly; a bound on hidden compute from power draw tightens sharply once the verifier can re-run
the declared work \citep{kimpson2026}, and in simulated datacentre power traces a tracker
with its threshold set on declared work detects training hidden in declared inference
\citep{kimpson2026datacentre}. The reference then plays the role of the template in warhead verification, where
the hard problem is to show that the template item is authentic \citep{yan2015}. A
classifier fitted elsewhere, such as the hardened forest of \citet{gargiulo2026}, might
screen recordings offered as declared work, but that forest also flags \GkHardPrefill\%
of our declared prefill (\cref{s:missed}).

\paragraph{What an alarm means.} An alarm from $s_R$ says only that the GPU ran work
unlike its declaration. That is, $s_R$ asks ``is this declared work?'', rather than ``is this
training?''. A higher score does not mean the work is more likely to be training. Hidden
training, honest training and non-AI work all score from \HidBandLo{} to \HidBandHi{}
(\cref{f:f7}), so the score cannot tell them apart. As a monitor it is in alarm for a fraction
\ARoneBusy{} of the time on undeclared non-AI work, where the classifier, $s_C^{\mathrm{dev}}$, stays silent
(\cref{t:t2}). Nor does an alarm show that a model was trained.
% results/ceiling/s2.json: tests.r1.primary, rate_a = rate_b = 1.0
The signature check flags every window of a run that discards every update, as it does
for training that applies its updates. An alarm therefore works like an anomaly under
nuclear safeguards~\citep[\S\S 11.24--11.26]{iaea2022glossary}. An alarm alerting does not necessarily mean that the operator has violated the agreement, but it triggers procedures for the verifier to investigate. The operator may have a legitimate reason for the departure, such as an unintentional error or a necessary deviation from the declared workload. The false-alarm rate of the detection method sets the expected frequency of such alerts, and the verifier must consider this when interpreting alarms. 
\paragraph{Limits.} This work has several limitations.
\begin{itemize}
  \item \emph{Scale.} We use one model of \ModelParams{}M parameters on one GPU. We do not test one probe on several GPUs at
        once. Frontier training runs across many GPUs and nodes; \citet{scher2025} envisage the signature check on clusters of
        many chips. We also look exclusively at current and do not consider other signals such as the communication between chips.
  \item \emph{Declared work.} Our declared work is the decode, prefill and idle of one
        model. An operator serving many models under varied load would declare far
        more, and every declared workload must be in the reference
        (\cref{s:declaration}). We do not test how the reference, or the cost of
        recording it, grows with the declaration.
  \item \emph{Threat model.} The evasions did not adapt to the detector. Our corpus lacks diluted low-rank
        adaptation, the evasion that most reduced detection in earlier studies
        (\cref{s:intro}). We test a corrupted reference only in the exploratory checks
        of \cref{s:monitoring}, with training drawn from our own runs rather than
        chosen by an adversary. \citet{scher2025}, by contrast, assume a well-resourced
        nation-state that can run arbitrary code to spoof chip activity.
  \item \emph{Measurement.} Our probe sits on the auxiliary supply and misses the
        current the GPU draws through its PCIe slot. The digitiser has no anti-alias filter (\cref{a:impl}). Our chain and
        that of \citet{gargiulo2026} also differ in sample rate, filtering and feature
        set (\cref{a:gk}).
  \item \emph{Calibration.} Our results depend on our split of the reference into a
        fit and a threshold set (\cref{t:t7}). One short calibration run each of decode and idle sets $\tau_\alpha$
        (\cref{t:corpus}), and resampling whole calibration runs moves $\tau_\alpha$
        over [\TauRoneLo, \TauRoneHi]. Calibration idle is also unlike test idle
        (\cref{s:declaration}).
  \item \emph{Exposure.} The test day holds no prefill and only \ExpDecodeH~h of
        declared decode, so the false-alarm bounds are loose, and declared prefill is a
        risk we could not measure on held-out data.
  \item \emph{Held-out data.} We measured detection of hidden training on development
        data before we scored the test day, so only the false-positive rates and the
        monitor results are held out.
\end{itemize}

\section*{Acknowledgements}
The measurements reported here were collected on an NVIDIA \GPU{} and current-sensing
hardware provided by \citet{amodo2026}. The author would like to thank
Carl Heimann of Amodo Design.

The author thanks the Machine Alignment, Transparency \& Security (MATS) programme---its
directors, organisers, funders, and staff---who made it possible to
work on this project and who provided invaluable resources and support
throughout.

\bibliography{references}
\bibliographystyle{icml2026}

\appendix
\crefalias{section}{appendix}
% Appendix tables and figures are numbered A1, A2, ... so a main-text citation shows where they are.
\setcounter{table}{0}
\renewcommand{\thetable}{A\arabic{table}}
\renewcommand{\theHtable}{A\arabic{table}}
\setcounter{figure}{0}
\renewcommand{\thefigure}{A\arabic{figure}}
\renewcommand{\theHfigure}{A\arabic{figure}}
\crefname{appendix}{Appendix}{Appendices}
\Crefname{appendix}{Appendix}{Appendices}
\onecolumn
\section{Implementation}\label{a:impl}

\paragraph{Channel.} A \Digitiser{} oscilloscope digitises the probe at \DigBits{}~bits
and streams the raw samples at \Fs{}~S/s. No filter precedes the converter, so current
variance above the Nyquist frequency folds into the band. We bound it from bursts
recorded at \BurstRate{}~MS/s, assuming no rejection. Under training the folded
variance is at most \AliasTrainPct\% of the in-band variance. Under idle, where the
current is small, it is at most \AliasIdleX{} times the in-band variance. At this bound,
both the raw-spectrum and the envelope features of idle windows would mostly describe
folded variance. Idle windows help set each detector's $\tau_\alpha$ and fit $s_R$, so a
different digitiser or an anti-alias filter could move them; we did not test this. The
low rate costs little signal. In the same bursts, \ChainBelowBand\% of a training trace's variance lies below
\ChainBandKhz{}~kHz, and the envelope at our rate matches one sampled ten times faster
to within \ChainEnvPct\%.

\paragraph{Corpus.} \Cref{t:corpus} counts runs, hours and windows per workload and
split.

\begin{table}[!htbp]
  \centering
  \small
  % generated by scripts/table_corpus.py from results/windows/manifest.json (repo_commit 018c3a2, corpus_commit 73ffebe)
% runs, scored hours and 5 s-hop windows per workload and split
\begin{tabular}{lrrrrrrrrr}
\toprule
 & \multicolumn{3}{c}{$\mathcal{D}_{\mathrm{dev}}$} & \multicolumn{3}{c}{$\mathcal{D}_{\mathrm{test}}$} & \multicolumn{3}{c}{$\mathcal{D}_{\mathrm{cal}}$} \\
\cmidrule(lr){2-4}\cmidrule(lr){5-7}\cmidrule(lr){8-10}
Workload & runs & h & windows & runs & h & windows & runs & h & windows \\
\midrule
honest training & 14 & 4.06 & 2905 & 4 & 0.95 & 680 & 2 & 0.48 & 340 \\
time-sharing & 3 & 0.71 & 510 & 7 & 2.22 & 1586 & -- & -- & -- \\
co-location & 3 & 0.72 & 510 & -- & -- & -- & -- & -- & -- \\
jitter & 4 & 0.95 & 678 & -- & -- & -- & -- & -- & -- \\
padding & 1 & 0.24 & 170 & -- & -- & -- & -- & -- & -- \\
power cap & 2 & 0.48 & 340 & -- & -- & -- & -- & -- & -- \\
declared decode & 2 & 0.48 & 340 & 2 & 0.48 & 340 & 1 & 0.24 & 170 \\
declared prefill & 1 & 0.24 & 170 & -- & -- & -- & -- & -- & -- \\
declared idle & 1 & 0.24 & 170 & 4 & 8.35 & 5994 & 1 & 0.24 & 170 \\
non-AI work & 4 & 0.95 & 681 & 4 & 0.95 & 680 & 2 & 0.48 & 340 \\
\midrule
total & 35 & 9.06 & 6474 & 21 & 12.94 & 9280 & 6 & 1.43 & 1020 \\
\bottomrule
\end{tabular}

  \caption{The corpus per workload and split: runs, scored hours and windows at the
    window-level hop. Time-sharing on the test day is the embedded episodes of
    \cref{s:works}.}
  \label{t:corpus}
\end{table}

\paragraph{Detectors.} The classifier's two stages are random forests and the
signature check is an isolation forest, all from scikit-learn with \ForestTrees{} trees,
default settings otherwise and a fixed seed.

\paragraph{Intervals.} For one run, a circular block bootstrap resamples blocks of
\BootBlock{} consecutive windows (\BootBlockS~s), \NBoot{} times
(\NBootBound{} for the episode lower bounds and the equivalence test). For two runs it
resamples within each run and pools. For three or more it resamples runs. Thresholds
stay fixed in every resample. On the monitor, the upper bound on $F$ assumes Poisson
onsets; the interval on $A$ resamples blocks of \OnsetBlockMin~min. An episode lower bound
is the smaller of a one-sided Clopper--Pearson bound and a block bootstrap over blocks
of \EpBlock{} consecutive episodes.

\paragraph{Episodes.} An episode that starts while an alarm is active counts as
detected only where the run's flag rate on gap windows, which hold no training, is at
most \GapGuard; we call this the gap guard. A run whose gaps hold no whole window takes the pooled gap flag rate
of the other runs. Otherwise only new onsets count, and the result is marked
onset-only. The delay $D$ is taken over episodes that raise a new alarm.
Under the complete reference, \EpPreRone{} of the \EpScored{} scored episodes start in alarm, and all are
credited (\cref{t:t7}).

\paragraph{Data and code.} The corpus is archived on Zenodo
\citep{kimpson2026corpus}. The analysis code, every committed result, and the scripts
that make every figure, table and number in this paper are at
\url{https://github.com/tomkimpson/verifying-declared-ai-workloads}. The window
features are not stored there; \texttt{scripts/make\_windows.py} regenerates them from
the archive. Every result
was computed against corpus commit \texttt{\CorpusCommit}.

\FloatBarrier
\section{The reference changes in detail}\label{a:reference}

\begin{figure}[!htbp]
  \centering
  \includegraphics[width=\linewidth]{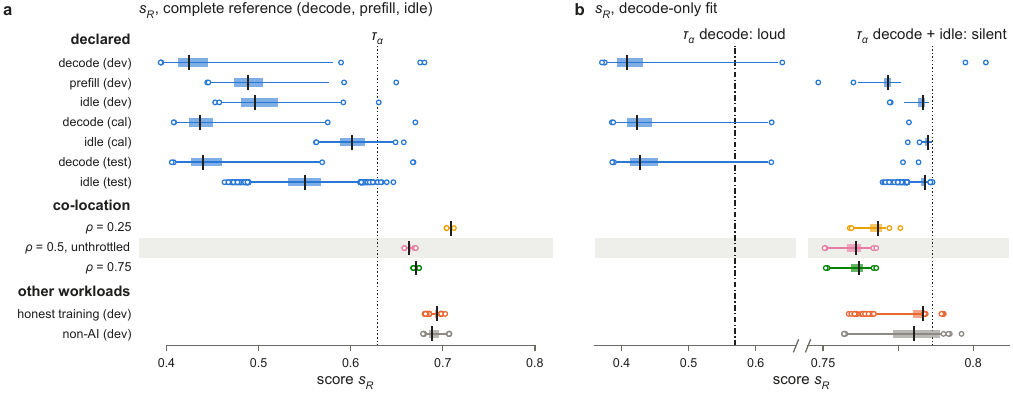}
  \caption{How $s_R$ scores each workload. Each row shows the per-window scores of one
    workload and split. Black ticks mark medians, bars the 25--75\% range and whiskers
    the 1--99\% range; circles are single windows beyond the whiskers. Windows to the right
    of $\tau_\alpha$ (dotted) are flagged. (a) The complete reference, with
    $\tau_\alpha = \TauRone$. It flags every co-location, honest-training and non-AI
    window. The shaded unthrottled run is the co-location run closest to $\tau_\alpha$; a
    few decode windows score above it (AUC \RoneAucDecode). (b) The decode-only fit
    (\cref{t:t7}), on a score axis broken at the slashes and widened beyond them. Prefill
    and idle are outside this fit, and idle now scores high. With idle in the threshold
    set, $\tau_\alpha = \TauDecOnlyPooled$ and no co-location window is flagged. With
    decode alone in it, $\tau_\alpha = \TauDecOnlyDecode$ and every co-location window
    and every idle window is flagged. Each threshold is labelled by its failure, as in
    \cref{t:t7}.}
  \label{f:f7}
\end{figure}

\paragraph{Scores.} \Cref{f:f7} shows the per-window scores behind the complete
reference and the decode-only fit of \cref{t:t7}.

\paragraph{Training in the threshold set.} We add $k$ honest-training windows to the
\EthreeCalWindows{} calibration windows, with the fit unchanged. Detection of the
unthrottled run falls to zero at $k = \EthreeKOne$ to \EthreeKTen{} for $\alpha$
from 1\% to 10\%, and at $k = \EthreeKFive$ for $\alpha = \AlphaPct\%$. At that $k$, honest
training is still flagged at $\ge \EthreeHonestMin\%$; it falls as $k$ grows further
(\cref{f:f9}a).

\paragraph{Training in the fit.} We add $k$ development honest-training windows, drawn at
random, to the \EfourNDeclared{} declared windows of the fit, with the threshold set
unchanged. At $k = \EfourKHonest$ (\EfourShareHonest\% of the fit), the calibration-day
full fine-tuning run, which is outside every fit, is flagged at \EfourFullCal\%, while
LoRA stays at \EfourLoraCal\%. The median detection of the unthrottled run is
\EfourHalfDet\% at $k = \EfourKHalf$ and \EfourGoneDet\% at $k = \EfourKGone$
(\EfourShareGone\% of the fit). The declared windows of the fit are flagged at
\EfourDeclZero\% with a clean fit, and at \EfourDeclEnd\% at the largest $k$ (median over
draws; \cref{f:f9}b). These windows are in every fit, so the rate is in-sample.

\section{The cross-check on the corpus of Gargiulo and Kulp}\label{a:gk}

We fetched their \GkNTraces{} traces \citep{gkdataset} at a pinned revision, checked each file's hash,
and resampled from \GkRate{}~MS/s to \GkDecimated{}~kS/s with an exact rational
resampler and its anti-alias filter. We extracted our \NFeat{} features and fitted
their two forests: genuine, on their training, inference and non-AI traces; hardened,
adding their adversarial traces.

On their own traces, under their grouped five-fold protocol, our measurement chain reaches
macro-$F_1$ \GkOwnFone{} (range \GkOwnFoneLo{} to \GkOwnFoneHi) and accuracy
\GkOwnAcc. They report \GkFone{} and \GkAcc{} with \GkNFeat{} features at the full
rate. We did not reach their accuracy, and scored our windows anyway. Shuffled labels give \GkShuffle. Their traces pass an anti-alias
filter on decimation and ours do not, so features above about \GkAliasKhz{}~kHz differ
between the datasets.

\paragraph{The published hardened forest.}
\citet{gargiulo2026} harden their forest on four evasion strategies of their own. We
scored our windows with both forests (\cref{t:t5}). At their default threshold of
\DefaultDecision, their genuine forest misses the
unthrottled run ($\GkGenMiss\%$). Their hardened forest catches all three
co-location runs, but it also flags \GkHardPrefill\% of declared prefill windows as
training. Their hardening works here. Its cost falls on a declared workload that our
calibration day does not hold.

\begin{table}[!htbp]
  \centering
  \small
  % generated by scripts/table_t5.py from results/gk/t5.json (repo_commit 4494926)
% percent of windows flagged training; co-location over windows with >= 1 complete step
\begin{tabular}{lrr}
\toprule
Our windows & G\&K hardened & G\&K genuine \\
\midrule
honest training ($\geq 80$\%) & 100.0 & 100.0 \\
declared decode & 0.0 & 0.0 \\
declared prefill & 100.0 & 2.4 \\
declared, decode + prefill ($\leq 20$\%) & 16.7 & 0.4 \\
\midrule
co-location, $\rho = 0.25$ & 100.0 & 100.0 \\
co-location, $\rho = 0.5$ & 100.0 & 0.0 \\
co-location, $\rho = 0.75$ & 100.0 & 100.0 \\
\bottomrule
\end{tabular}

  \caption{Forests fitted by the recipe of \citet{gargiulo2026} (G\&K) on their corpus,
    scored on our windows: per cent flagged training. The first rows are a
    transfer gate (training $\geq$ \GateTrain\%, declared
    $\leq$ \GateInf\%). Both forests pass. The hardened forest passes the declared
    row only because decode outnumbers prefill. Their chain and ours differ in rate
    and anti-alias filtering, and their own-data check fell short of their reported
    accuracy (\cref{a:gk}).}
  \label{t:t5}
\end{table}

\FloatBarrier

\end{document}